\documentclass[12pt,preprint]{aastex}

\newcommand{\pfrac}[2]{\left(\frac{#1}{#2}\right)}
\newcommand{\mfrac}[2]{\left[\frac{#1}{#2}\right]}

\def\Mesz{M\'esz\'aros~}

\def\beq{\begin{equation}}
\def\enq{\end{equation}}
\def\bea{\begin{eqnarray}}
\def\ena{\end{eqnarray}}
\def\bec{\begin{center}}
\def\enc{\end{center}}
\def\etal{{et al.~}}
\def\eps{\epsilon}
\def\ob{{\rm ob}}
\def\pair{$e^{\pm}~$}

\shorttitle{Prompt UV-to-soft-X-ray emission of GRBs}
\shortauthors{Li \& Song}

\begin{document}

\title{Prompt Ultraviolet-to-Soft-X-Ray Emission of Gamma-Ray Bursts:
Application to GRB 031203?}

\author{Zhuo Li\altaffilmark{1} and L. M. Song\altaffilmark{1}}
\altaffiltext{1}{Particle Astrophysics Center, Institute of High
Energy Physics, Chinese Academy of Sciences, Beijing 100039,
China}

\begin{abstract}
We discuss the prompt emission of GRBs, allowing for
$\gamma\gamma$ pair production and synchrotron self-absorption.
The observed hard spectra suggest heavy pair-loading in GRBs. The
re-emission of the generated pairs results in the energy
transmission from high-energy gamma-rays to long-wavelength
radiation. Due to strong self-absorption, the synchrotron
radiation by pairs is in optically thick regime, showing a
thermal-like spectral bump in the extreme-ultraviolet/soft X-ray
band, other than the peak from the main burst. Recently, the
prompt soft X-ray emission of GRB 031203 was detected thanks to
the discovery of a delayed dust echo, and it seems to be
consistent with the model prediction of a double-peak structure.
The confirmation of the thermal-like feature and the double-peak
structure by observation would indicate that the dominant
radiation mechanism in GRBs is synchrotron rather than
inverse-Compton radiation.
\end{abstract}


\section{Introduction}
In the past few years, a standard model was well established in
which the gamma-ray burst (GRB) afterglows result from the
relativistic blast-waves sweeping up the ambient medium of GRBs
(\Mesz 2002). However, the prompt emission of GRBs is believed to
be irrelevant to ambient medium, and its radiation mechanism is
still poorly known so far.

The recent definite proof of GRB 030329 associated with a type
Ib/c supernova confirmed, as long suspected, that GRBs, at less
the long class, originate from explosions of massive stars in
distant galaxies (Stanek \etal 2003; Hjorth \etal 2003). Since
GRBs are events occurring on stars, the emission region may be
compact, and the huge energy release will lead to the formation of
$e^\pm,\gamma$ fireballs, exhibiting thermal-like spectra. But the
GRB spectra are non-thermal and hard, with a significant fraction
of the energy above the \pair pair formation energy threshold. For
a photon with tens of MeV to escape freely, avoiding
$\gamma\gamma$ interactions, the fireball must be
ultra-relativistic expanding, with $\Gamma\ga100$ (Lithwick \&
Sari 2001, and references therein). The afterglow studies has also
confirmed the presence of ultra-relativistic motion. However, if
the intrinsic emission, before leaking out from fireball, includes
radiation with even higher energy, say, beyond GeV, these
radiation still suffers $\gamma\gamma$ absorption, leading to pair
loading in GRBs. In context of relativistic fireball model, Li
\etal (2003; Li03 hereafter) found that, in a wide range of model
parameters, the resulting pairs may dominate those electrons
associated with fireball baryons. The presence of abundant pairs
would affect the behaviors of the early afterglow from reverse
shocks (Li03), and may also emit particular signals in the
bursting phase.

We discuss in this Letter the prompt GRB emission, with emphasis
on the re-emission by the secondary \pair pairs. If the energy
density in the emission region is dominated by magnetic field, the
pairs would re-emit mainly by synchrotron radiation, rather than
IC process (e.g., Pilla \& Loeb 1998). Due to strong
self-absorption, the pair emission appears as a thermal-like bump
in the GRB spectrum, similar to the feature discussed by
Kobayashi, \Mesz \& Zhang (2004) in the context of reverse shock
emission. (Fan \& Wei 2004 have also studied the pair emission,
but with less stress on the self-absorption effect.) We further
show that the intense soft X-ray emission in GRB 031203, inferred
by the delayed dust halo (Vaughan \etal 2004 [V04]; Watson \etal
2004 [W04]), can be accounted for by the spectral bump due to
pair-loading. This is of significant interest, since this feature
could give a diagnostics for the magnetic field in the fireball
(Kobayashi, \Mesz \& Zhang 2004) and the dominant radiation
mechanism in GRBs.

\section{Pair loading in GRB fireballs}
Consider a GRB central engine that produce a relativistic wind
outflow, with an isotropic energy $E$, a bulk Lorentz factor
$\Gamma$ and a width $\Delta$. The energy carried in the outflow
may be composed of two components, the bulk kinetic energy of
baryons ($E_k$) and the energy of magnetic field ($E_B$). The
ratio between them can be defined as $\sigma\equiv E_B/E_k$ (e.g.,
Zhang \& \Mesz 2002). These energies are carried from the central
engine to some radius $R$ where GRB emission arises. As in Li03,
the emission site can be constrained by the non-thermal spectra
and rapid varying light curves of GRBs, leading to typical value
of $R\sim10^{14}$~cm for $\Gamma\sim300$. The width of outflow is
$\Delta\la 10^{12}$~cm for a wind lasting a duration of
$T\la100$~s. Thus the emission region can be regarded as a thin
shell with $\Delta\ll R$.

In the kinetic-energy dominated model ($\sigma<1$), the bulk
kinetic energy is dissipated by internal shock waves within the
unsteady outflow, where the magnetic field strength $B$ is in the
equipartition value $B\sim(8\pi
U_\gamma)^{1/2}=10^4L_{\gamma,51}^{1/2}\Gamma_{300}^{-1}R_{14}^{-1}$~G,
with $L_\gamma=10^{51}L_{\gamma,51}$~erg s$^{-1}$ the GRB
luminosity, $\Gamma_{300}=\Gamma/300$ and $R_{14}=R/10^{14}$~cm.
In the magnetic-energy dominated model, the magnetic field could
be stronger than the equipartition value, leading to a small
radiation-to-magnetic energy ratio in emission region, $Y\equiv
U_\gamma/U_B<1$. Though broad band fits of afterglows generally
give $Y>1$ for shocked-medium, the presence of highly magnetized
ejecta is suggested by some recent works (e.g., Zhang, Kobayashi,
\& \Mesz 2003). Here we will assume $Y<1$ for GRBs, and scale the
magnetic field as $B=10^4B_4$~G.

Due to the large luminosity and hard spectrum of a GRB, intrinsic
high energy gamma-rays produced in the GRB emission region could
be absorbed for pair production. As in Li03, the cut-off energy,
above which the photons suffer strong absorption, and the number
of produced pairs can be estimated from the observed GRB spectra.
The observed photon spectra of GRBs can be approximated by a
broken power-law, with a high-energy portion of the form
$dN_\gamma/d\eps\propto\eps^{-\beta}$ for $\eps>\eps_p$, where
$\eps_p\sim m_ec^2$ is the energy at the broken point and the
index $\beta\sim2-3$. The number of the produced secondary pairs
is equal to the absorbed photons above $\eps_{cut}$. Assuming the
intrinsic spectrum above $\eps_{cut}$ follows the same power law
below $\eps_{cut}$, we calculate the pair number as $
N_\pm=N_\gamma(>\eps_{cut})
\simeq(E_\gamma/\eps_p)(\eps_{cut}/\eps_p)^{-(\beta-1)}$. Since
the timescale of $\gamma\gamma$ collisions (comoving frame),
$t'_{\gamma\gamma}\simeq[(\sigma_T/5)n_\gamma'
c]^{-1}=0.2\Gamma_{300}R_{14}^2L_{\gamma,51}^{-1}$, is usually
shorter than the dynamical time (comoving frame), $t'_{dyn}\simeq
R/\Gamma c=10R_{14}\Gamma_{300}^{-1}$~s, the resulting pairs
remain inside the outflow.

As in Li03, $\eps_{cut}$ should be defined by the photon energy at
which the optical depth for $\gamma\gamma$ absorption equals
unity, $\tau_{\gamma\gamma}(\eps)=1$, where the optical depth can
be given by a simplified expression under the thin-shell
assumption of the emission region,
$\tau_{\gamma\gamma}(\eps)=(11/180)\sigma_TN_\gamma(>\eps)/4\pi
R^2$ (Lithwick \& Sari 2001). Furthermore, the observed cut-off
energy must be larger than $\Gamma m_ec^2$. In summary,
\begin{equation}
\eps_{cut}=\max\left[0.3\pfrac{R_{14}^2}{E_{\gamma,52}\eps_0^{\beta-2}}^{1/(\beta-1)}\Gamma_{300}^2;
~~0.2\Gamma_{300}\right]{\rm ~GeV},
\end{equation}
where $\eps_0=\eps_p/m_ec^2$, and hereinafter the numerical
coefficient corresponds to $\beta=2.4$. It can be seen that the
detection of the cut-off energy can help to constrain $\Gamma$ and
$R$. EGRET had detected prompt GeV emission, without obvious
attenuation, in several GRBs (e.g., GRB 930131; Sommer \etal
1994). We expect that the future satellite {\it GLAST}, which
works in 10 MeV$-$300 GeV range, could observe such a cut off at
multi-GeV. \footnote{Cosmic infrared background can also absorb
high energy photons, but primarily in TeV range (Salamon \&
Stecker 1998).} With $\eps_{\rm GeV}=\eps_{cut}/1$~GeV, the pair
number is written as
\begin{equation}\label{Npm}
N_\pm\simeq 3\times10^{53}E_{\gamma,52}\eps_{\rm
GeV}^{-(\beta-1)}\eps_0^{\beta-2}.
\end{equation}
For comparison, the number of baryonic electrons in the fireball
is $N_b=E/(1+\sigma)\Gamma
m_pc^2=2\times10^{52}E_{52}\Gamma_{300}^{-1}(1+\sigma)^{-1}$, with
$E_{52}=E/10^{52}$~erg. So pairs become the dominant component.
The baryonic electrons are expected to be responsible for the
prompt hard X-ray emission, whilst the pairs might give rise to
low energy emission, discussed below.

\section{Extreme-ultraviolet bump in the prompt emission}
Since the energy of a generated $e^+$ ($e^-$) comes primarily from
the photon with lager energy between the two colliding ones, the
initial energy distribution of the generated pairs would follow
the form of the high-energy spectral tail, i.e.,
$dn_\pm/d\gamma_e\propto\gamma_e^{-\beta}$ for
$\gamma_e>\gamma_\pm$, where $\gamma_\pm$ corresponds to the
cut-off energy, $\gamma_\pm=\eps_{cut}/2\Gamma
m_ec^2\simeq3.3\eps_{\rm GeV}\Gamma_{300}^{-1}$. These pairs will
cool down by synchrotron rather than IC radiation in the $Y\la1$
condition here. Because of the strong magnetic field in the
emission region, the synchrotron-cooling timescale of pairs,
$t'_{syn}=8B_4^{-2}\gamma_\pm^{-1}$~s, is shorter than the
fireball dynamical time, implying that the pairs are always fast
cooling. We assume that pair annihilation is negligible, confirmed
later.

For these fast cooling pairs, their energies are emitted quickly.
As a result, the energy above $\eps_{cut}$ in the intrinsic
spectrum re-arises as the pair emission. The luminosity of the
pair emission is given by
\begin{equation}\label{Lpm}
L_\pm\simeq\frac{\beta-1}{\beta-2}
\frac{N_\pm\eps_{cut}}T\simeq2\times10^{50}L_{\gamma,51}\pfrac{\eps_0}{\eps_{\rm
GeV}}^{\beta-2}{\rm ~erg\,s}^{-1},
\end{equation}
with $L_\gamma=E_\gamma/T=10^{51}L_{\gamma,51}$~erg s$^{-1}$ the
GRB luminosity, and the characteristic synchrotron frequency is
\begin{equation}
\nu_\pm=0.9\times10^{14}\Gamma_{300}^{-1}\eps_{\rm GeV}^2B_4{\rm
~Hz}.
\end{equation}
If we assume $Y\ll1$, the synchrotron radiation plays a dominant
role of pair cooling, rather than IC process. In this condition,
the luminosity $L_\pm$ will peak at frequency $\nu_\pm$. Thus a
very intense optical flash will emerge accompanying the prompt
gamma-rays if neglecting the self-absorption. However, as shown in
the following, the self-absorption is strong in such low energy
range, with the absorption frequency $\nu_a\gg\nu_\pm$, i.e., most
of the pair emission occurs in the optically thick regime. Similar
to the case of reverse flash in the condition of
$\nu_c<\nu_m<\nu_a$, which is discussed by Kobayashi \etal (2004),
a thermal-like bump will arise in the low energy range of a GRB
spectrum.

The self-absorption suppresses the emission below absorption
frequency $\nu_a$, and the suppressed emission energy is
redistributed again among the pairs, preventing the pairs cool
down immediately. So, the pairs and the radiation obtain a
mechanism to exchange their energies. The final result is that the
initial injected pair energy is redistributed among pairs and
radiation, leading to a bump in the spectrum. The emission in the
hard X-ray band is not in the optically thick regime, and is not
involved in the energy redistribution. In the GRB duration $T$,
the pair energy is radiated around $\nu_a$, where the flux is
given by $F_{\nu_a}\simeq L_\pm/4\pi D_L^2\nu_a$, with $D_L$ the
GRB luminosity distance. We follow the simple way by Sari \& Piran
(1999)
to estimate the maximal flux as a blackbody with the pair
temperature, $
F_{\nu_a,bb}\approx\pi(R_\perp/D_L)^2(2\nu_a^2/c^2)kT_\pm$, where
$R_\perp\simeq R/\Gamma$ is the observed size of the fireball, the
pair temperature is $kT_\pm\simeq\Gamma\gamma_am_ec^2/3$, and
$\gamma_a$ is the pair Lorentz factor that corresponds to $\nu_a$
and is given by $(2\pi m_ec\nu_a/\Gamma eB)^{1/2}$. Equating
$F_{\nu_a,bb}\simeq F_{\nu_a}$ yields the self-absorption
frequency
\begin{equation}\label{nua}
\nu_a\simeq1\times10^{16}L_{\pm,50}^{2/7}\Gamma_{300}^{3/7}R_{14}^{-4/7}B_4^{1/7}{\rm
Hz},
\end{equation}
which is in the extreme-ultraviolet (EUV) band. Since
$\nu_a\gg\nu_\pm$, most of the emission is absorbed and
re-distributed, giving rise to a black-body like bump in the GRB
spectrum, with peak frequency around $\nu_a$ (eq.[\ref{nua}]) and
luminosity $L_\pm$ (eq.[\ref{Lpm}]).

The comoving-frame annihilation timescale of the pairs in the
thermal-like bump, with the comoving-frame temperature
$\gamma_am_ec^2/3$, is $t_{ann}'\simeq(\sigma_{ann}n_\pm c)^{-1}$,
where $\sigma_{ann}\simeq(3\sigma_T/8\gamma_a)(\ln2\gamma_a-1)$ is
the annihilation cross section, and the pair number density is
given by $n_\pm\approx n_{\gamma}(>\eps_{cut})\simeq L_\pm/4\pi
R^2\Gamma c\eps_{cut}$. Note the baryonic electrons are neglected.
The annihilation fraction of pairs, $f_{ann}$, can be estimated by
$t_{ann}'^{-1}$ times the comoving-frame dynamical timescale
$t_{dyn}'$,
$f_{ann}\sim0.08(L_{\pm,51}/\Gamma_{300}^3R_{14})(\gamma_\pm\gamma_a/10)^{-1}$.
Since $f_{ann}\ll1$, the annihilation in the bump is really
negligible.

Notice that $\nu_a$ is insensitive to all the parameters (eq.
\ref{nua}), and would be fixed in the EUV range for various GRBs.
However, as it propagates, the EUV radiation from a GRB is subject
to intergalactic or galactic absorption (e.g., Gou \etal 2004),
hence only the optical/UV or soft X-ray emission is expected for
observation. Below $\nu_a$ the spectrum behaves as
$F_{\nu<\nu_a}=F_{\nu_a}(\nu/\nu_a)^2$, then the observed pair
emission at 1 eV is
\begin{equation}\label{UVflux}
F_{\nu}^\ob({\rm 1 ~eV})
\simeq0.9\frac{L_{\gamma,51}(1+z)^3}{D_{28}^{2}\nu_{a,16}^{3}}\pfrac{\eps_0}{\eps_{\rm
GeV}}^{\beta-2}{\rm mJy},
\end{equation}
where $\nu_{a,16}=\nu_a/10^{16}$~Hz, and we have obviously shown
the dependence on redshift $z$. Whereas in the band above $\nu_a$,
the emission is in optically thin regime and still exhibits the
form radiated by the initial pairs,
$F_{\nu>\nu_a}=F_{\nu_\pm}(\nu/\nu_\pm)^{-\beta/2}$, where
$F_{\nu_\pm}\simeq L_\pm/4\pi D_L^2\nu_\pm$. If observed at 1 keV,
the flux contributed by pairs is then
\begin{eqnarray}\label{Xflux}
F_{\nu}^\ob({\rm 1~keV})\simeq
3\times10^{-8}\frac{L_{\gamma,51}}{D_{28}^{2}}\mfrac{\eps_0^2B_4}{\Gamma_{300}(1+z)}^{(\beta-2)/2}\nonumber\\
{\rm erg}\,{\rm cm}^{-2}{\rm s}^{-1}{\rm keV}^{-1}.
\end{eqnarray}
This calculation is valid until at high enough frequency where the
main GRB peak dominates the emission. For the typical parameters,
the contrast of the thermal bump relative to the power-law
spectrum $(\nu_a/\nu_\pm)^{\beta/2-1}$ (Fig. 1) is a factor of a
few, and the bump is weak. For larger index $\beta\sim3$ the
contrast is larger, and though the energy injected into the pairs
becomes smaller, the thermal bump still sticks out above the
primary emission component. The prompt optical/UV
(eq.\ref{UVflux}) and X-ray (eq.\ref{Xflux}) emission are expected
to be observed by the UVOT and XRT detector, respectively, on
board the up-coming {\it Swift} satellite.

As a result, the intrinsic high-energy emission in the GRB
spectrum is absorbed and then transferred to a thermal-like bump
in the EUV band, as shown in Fig. 1. Since these features are
expected to arise if $Y\ll1$, the observation would provide a
constraint on the magnetization parameter and the radiation
mechanism in GRBs.

\section{GRB 031203}
GRB 031203 was detected by {\it INTEGRAL} as a single-pulse burst
with a duration of 30 s and a peak flux of $1.3\times10^{-7}$ erg
s$^{-1}$~cm$^{-2}$ in the 20-200 keV band (G\"otz \etal 2003;
Mereghetti \& G\"otz 2003). A double exponential approximation to
the single-pulse light curve yields an estimated fluence of
$4\times10^{-7}$erg cm$^{-2}$ (20-200~keV) (Prochaska \etal 2004;
P04). With the redshift $z=0.1055$ measured from the optical
observation of the host galaxy, the $k$-correction
isotropic-equivalent energy release is estimated to be $E_{\rm
iso}(20-2000~{\rm keV})=2.6\times10^{49}$erg ($h=0.7$,
$\Omega_\Lambda=0.7$ and $\Omega_m=0.3$) (P04).

6 hours after the burst, {\it XMM-Newton} discovers a
time-dependent, dust-scattered X-ray halo around the burst (V04;
W04). The halo brightness implies an initial soft X-ray pulse
consistent with the burst. Since few satellites observe GRBs in
soft X-ray band, this observation is important for us to know the
emission features of GRBs in such low energy range. Based on the
hypothesis of the column of dust along the sightline toward GRB
031203, V04 and W04 estimates a source flux of
$1.5\pm0.8\times10^{-7}$erg cm$^{-2}$s$^{-1}$ with photon index
$\beta_X=2.2\pm0.3$ in the 0.2-10 keV band. This leads them to
regard GRB 031203 as an X-ray flash with peak energy at
$\la10$~keV. However, P04 argue that the above analysis has
overestimated the scattering column of dust, which may be 4.4 (or
even 27) times larger, and hence the source should be 4.4 (27)
times smaller in flux. Thus, we take in the following discussion
the source flux of the dust halo as
$3.4\pm1.8~(0.56\pm0.30)\times10^{-8}$erg cm$^{-2}$s$^{-1}$ in the
0.2-10 keV band, which corresponds to a fluence of
$10\pm5.4~(1.7\pm0.9)\times10^{-7}$erg cm$^{-2}$ for a duration of
30~s.

The observational results of GRB 031203 is consistent with the
double-peak spectral shape described in \S 3. This rests on two
points. First, the photon index of $\sim2.2$ in 0.2-10 keV band
implies a soft X-ray peak, with most energy released at
$\la0.2$~keV. And second, the fluence in 20-200 keV band is larger
than that extrapolated from the 0.2-10 keV band using the photon
index $\beta_X\sim2.2$, suggesting a second peak in the hard X-ray
band. The observed hard X-ray emission is only comparable to or
even slightly smaller than the soft X-ray one in fluence, but we
expect that, since the redshift of GRB 031203 is low and hence the
peak energy is less redshifted, the hard X-ray emission could peak
at higher energy, $\eps_p>200$~keV. If peaking at $\sim1$~MeV, the
hard X-ray fluence would be larger by a factor of $\sim5$, and
dominates the soft one. It is unfortunate that there is not
reliable time-integrated spectrum of GRB 031203 above 20 keV that
is given. Therefore we cannot compare the two peaks for more
details. Since the soft X-ray spectrum of GRB 031203 follows a
power law, the thermal-like bump should be at an energy below the
observation window, $\la0.2$~keV, consistent with predicted in
eq.(\ref{nua}).

There may be some caveats here. One may think that a double-peak
structure can also be interpreted as synchrotron self-Compton
radiation provided $Y>1$. But in this condition, the pairs from
$\gamma\gamma$ production will loss energy by IC radiation and
result in a spectral bump at $\gamma_\pm^2\eps_p\la100$~MeV,
something conflicting with the observation of GRBs in high energy
band (Schaefer \etal 1998). A further difficulty for SSC to
interpret the double peaks is that there may be many Compton order
if $Y>1$, and each higher Compton order will dominate over the
previous one by the same amount $Y$ until the typical emitted
energy reaches the electron energy. Only a small fraction of the
radiated power would therefore be observed in the sub-MeV band
(e.g., Ghisellini \etal 2000), resulting in energy crisis of GRBs.

One may also think that the source of the dust halo is the X-ray
afterglow in early time rather than the prompt burst. We believe
this can be ruled out. The dust halo consists of two narrow rings,
and V04 have interpret them as being due to the prompt burst
scattered by two distinct scattering screens. One may still
propose that only a single scattering screen is required as long
as there is a complex early time structure of the X-ray afterglow.
However, if so the later ring should evolve with a certain time
delay with respect to the first one, which is in contrast with the
observation (see $\theta-t$ curves in fig. 3 of V04). The ring is
rather narrow, implying that the intrinsic emission is a short
pulse, as opposed to the smooth afterglow behavior. In fact, since
the ring is narrow, $\Delta\theta\sim20$~arcsec in angular width,
the pulse duration is limited to $\Delta t\la
D\Delta\theta^2/c\sim800$~s, where the distances of the screens
from earth are $D\sim1$~kpc for both rings (V04). Because of
causality, the dynamical time is also limited to $t\sim\Delta
t\la800$~s.

\section{Summary and discussion}
We have studied the prompt GRB emission, allowing for
$\gamma\gamma$ pair production and synchrotron self-absorption.
Inferred by the observed characteristics of GRB emission, the
resulting pairs usually dominate the baryonic electrons. The pairs
will give rise to further emission by synchrotron radiation if in
the strong magnetic field, which is also responsible to the prompt
sub-MeV emission. However, due to strong self-absorption the pair
emission exhibits a thermal-like bump in the extreme UV/soft X-ray
band, other than the peak in the hard X-ray band. Since the
feature emerges for $Y<1$, its observation gives a diagnostics for
the magnetic energy density in the fireball (Kobayashi, \Mesz, \&
Zhang 2004). The recent observation of a dust halo around GRB
031203 infers a spectral peak of the prompt burst emission in the
soft X-ray band, which seems to be consistent with the predicted
double-peak structure.

Some primary hypotheses have been taken in our calculation. First,
we assume that the emission region is transparent for Compton
scattering, even though the secondary pairs increase significantly
the total optical depth. For typical parameter values this
assumption is protected. However, in some extreme cases with quite
small $\Gamma$ and $R$, the secondary pairs may form an optically
thick screen again (Guetta, Spada \& Waxman 2001; Kobayashi, Ryde
\& MacFadyen 2002), which degrades the gamma-rays and results in
an X-ray flash (XRF; \Mesz \etal 2002). If so, our calculation
using eq. (\ref{Npm}) may underestimate the pair-loading in XRFs,
which may need detailed works of numerical simulation (e.g., Pe'er
\& Waxman 2003). Secondly, we assume strong magnetic field, $Y<1$,
in the emission region. If $Y>1$, the pairs lose most energy by IC
scattering the GRB photons, and the IC photons are not
self-absorbed again since beyond the optically thick regime, hence
no effective energy exchange between pairs and photons is
established and the bump disappears. Therefore, once UV/soft X-ray
bumps are detected this will infer $Y<1$ and that it is
synchrotron rather than IC radiation that gives rise to the
sub-MeV emission of GRBs.

\acknowledgments

This work was supported by the National 973 Project and the
Special Funds for Major State Basic Research Projects.

\begin{figure}
\epsscale{.90} \plotone{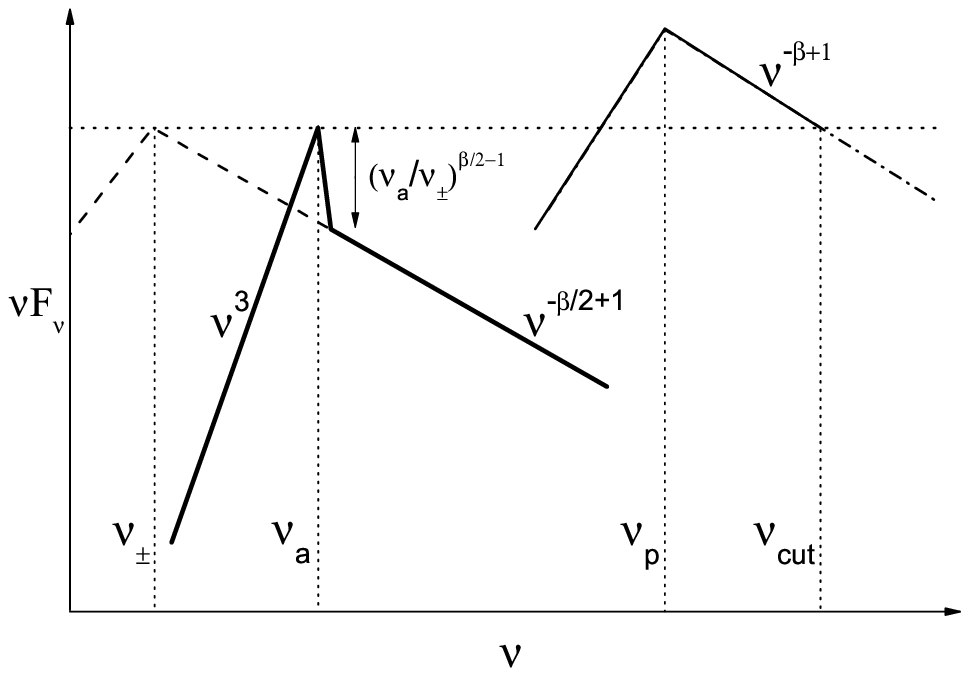} \caption{The schematic diagram of
the $\nu F_\nu$ spectrum of the GRB prompt emission. Note that
$\nu_p=\eps_p/h$ and $\nu_{cut}=\eps_{cut}/h$. The {\it thin
solid} line shows the GRB emission with cut-off above $\nu_{cut}$
due to $\gamma\gamma$ pair production, while the {\it
dashed-dotted} line is the intrinsic spectrum without cut-off. The
synchrotron radiation by the resulting pairs is shown by the {\it
thick solid} or {\it dashed} lines, corresponding to with or
without self-absorption, respectively. The maximum around $\nu_a$
would appear as a thermal-like peak.}
\end{figure}

\end{document}